\documentclass[conference]{IEEEtran}
\IEEEoverridecommandlockouts
\usepackage{cite}
\usepackage{amsmath,amssymb,amsfonts}
\usepackage{algorithmic}
\usepackage{siunitx}
\usepackage{graphicx}
\usepackage{textcomp}
\usepackage{xcolor}
\usepackage[a4paper, total={184mm,239mm}]{geometry}
\def\BibTeX{{\rm B\kern-.05em{\sc i\kern-.025em b}\kern-.08em
        T\kern-.1667em\lower.7ex\hbox{E}\kern-.125emX}}

\newcommand{\faultyVP}{\emph{X-Fault}}

\usepackage{fancyhdr}
\usepackage[all]{background}
\usepackage{stackengine}
\setstackEOL{\\}
\setstackgap{L}{\normalbaselineskip}
\SetBgContents{\color{gray}{\tiny \Longstack{PREPRINT - accepted In Proceedings of the 4th International Conference on Artificial Intelligence Circuits and Systems (AICAS '22)}}}% Set contents
\SetBgPosition{4.2cm,1cm}% Select location
\SetBgOpacity{1.0}% Select opacity
\SetBgAngle{0}% Select rotation of logo
\SetBgScale{1.8}% Select scale factor of logo
\usepackage{pgfplots}
\usepackage{pgfplotstable}
\usetikzlibrary{patterns, arrows, pgfplots.groupplots,hobby}
\usetikzlibrary{patterns}
\usetikzlibrary{positioning}
\usetikzlibrary{pgfplots.groupplots}
\usetikzlibrary{plotmarks}
\usetikzlibrary{calc}
\usepgfplotslibrary{external}
\usepackage{balance}
\usepackage{caption}
\usepackage{xcolor}
\usepackage{colortbl}
\usepackage{rotating}
\usepackage{multirow}
\usepackage{booktabs}
\usepackage{bm}
\usepackage{subfig}
\usepackage{enumitem}
\usepackage{soul}
\usepackage[export]{adjustbox}
\usepackage{bigstrut}
\definecolor{cg1}{HTML}{E8F1FA}
\definecolor{cg2}{HTML}{C7DDF2}
\definecolor{cg3}{HTML}{8EBAE5}
\definecolor{cg4}{HTML}{407FB7}
\definecolor{cg5}{HTML}{00549F}
\definecolor{red}{HTML}{CC071E}

\definecolor{cg1}{HTML}{9C6261}
\definecolor{cg2}{HTML}{E98667}
\definecolor{cg3}{HTML}{E8B78E}
\definecolor{cg4}{HTML}{DE9853}
\definecolor{cg5}{HTML}{775445}
\definecolor{cg6}{HTML}{3C3D28}

\usepackage{keyval}

\begin{document}
\bstctlcite{IEEEexample:BSTcontrol}

\title{\faultyVP: Impact of Faults on Binary Neural Networks in Memristor-Crossbar Arrays with Logic-in-Memory Computation
\thanks{This work was funded by the Federal Ministry of Education and Research (BMBF, Germany) in the project NEUROTEC II (16ME0398K, 16ME0399).}
%\vspace{-4mm}
}

\author{
    \IEEEauthorblockN{
    Felix Staudigl\IEEEauthorrefmark{1},
    Karl J. X. Sturm\IEEEauthorrefmark{1},
    Maximilian Bartel\IEEEauthorrefmark{1},
    Thorben Fetz\IEEEauthorrefmark{1},\\
    Dominik Sisejkovic\IEEEauthorrefmark{1},
    Jan Moritz Joseph\IEEEauthorrefmark{1},
    Leticia Bolzani P\"ohls\IEEEauthorrefmark{2},
    and Rainer Leupers\IEEEauthorrefmark{1}
}
%  \vspace{+2mm}
    \IEEEauthorblockA{\IEEEauthorrefmark{1}
        \textit{Institute for Communication Technologies and Embedded Systems, RWTH Aachen University, Germany}
    }
    \IEEEauthorblockA{\IEEEauthorrefmark{2}
        \textit{Chair of Integrated Digital Systems and Circuit Design, RWTH Aachen University, Germany}\\
    \{staudigl, sturm, bartel, fetz, sisejkovic, joseph, leupers\}@ice.rwth-aachen.de\\
    poehls@ids.rwth-aachen.de\\
}
\vspace{-9mm}
%\vspace{20mm}
}

\maketitle

\begin{abstract}
Memristor-based crossbar arrays represent a promising emerging memory technology to replace conventional memories by offering a high density and enabling computing-in-memory (CIM) paradigms. While analog computing provides the best performance, non-idealities and ADC/DAC conversion limit memristor-based CIM. Logic-in-Memory (LIM) presents another flavor of CIM, in which the memristors are used in a binary manner to implement logic gates. Since binary neural networks (BNNs) use binary logic gates as the dominant operation, they can benefit from the massively parallel execution of binary operations and better resilience to variations of the memristors. Although conventional neural networks have been thoroughly investigated, the impact of faults on memristor-based BNNs remains unclear. Therefore, we analyze the impact of faults on logic gates in memristor-based crossbar arrays for BNNs. We propose a simulation framework that simulates different traditional faults to examine the accuracy loss of BNNs on memristive crossbar arrays. In addition, we compare different logic families based on the robustness and feasibility to accelerate AI applications.
\end{abstract}

\begin{IEEEkeywords}
ReRAM, memristor, faults, reliability, logic-in-memory\vspace{-3mm}
\end{IEEEkeywords}

\section{Introduction}

Non-volatile memories (NVMs) such as resistive RAM (ReRAM) offer advantages over conventional RAMs in terms of density, power consumption, and computing-in-memory (CIM) capabilities. CIM is most promising to address the von Neumann bottleneck by reducing data exchange, thus addressing the limitations of memory-bound AI accelerators~\cite{Staudigl_2021,joseph}.

\begin{figure}[ht]
    \centering
    \includegraphics[width=\columnwidth]{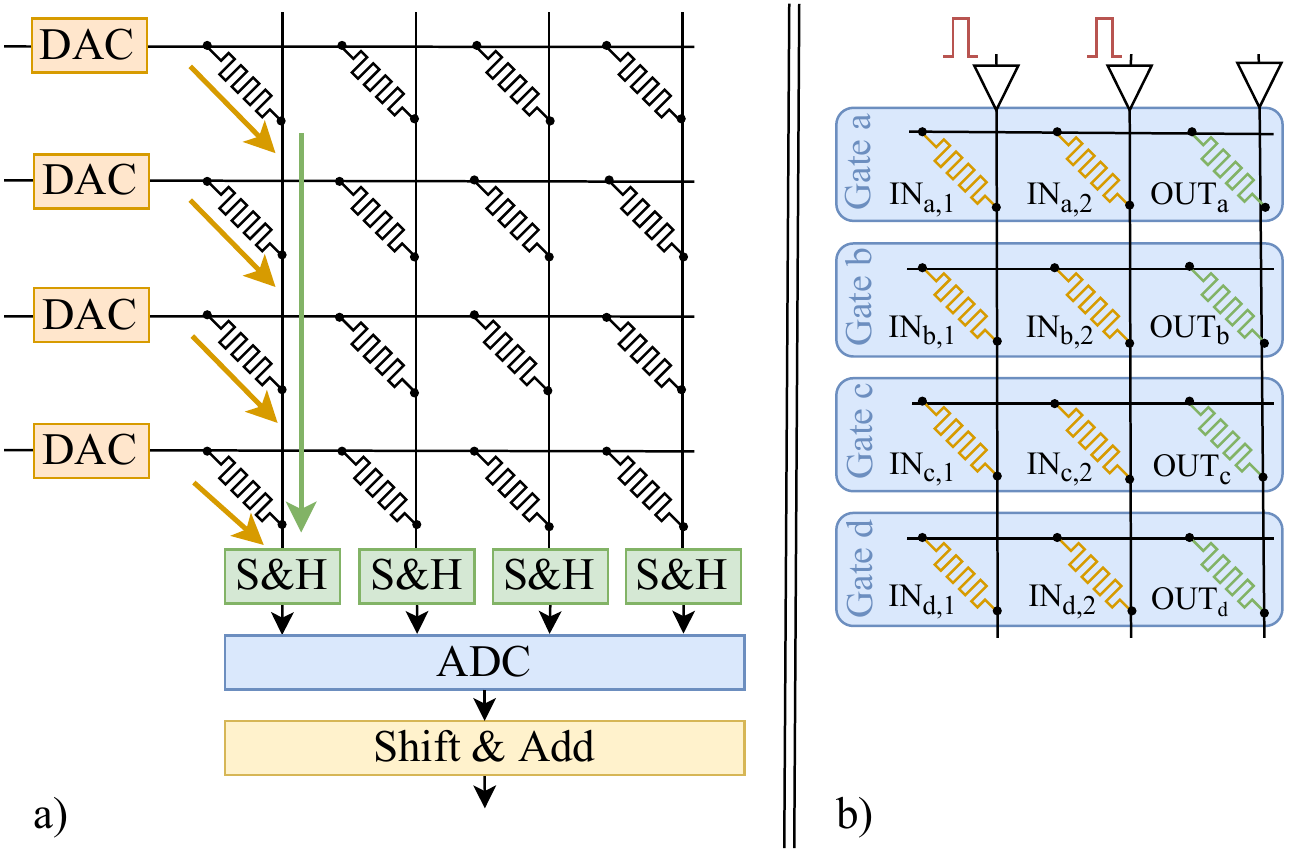}
    \caption{Comparison of CIM paradigms: (a) Kirchhoff-based analog CIM, and (b) binary logic-in-memory.}
    \label{fig:intro}
    \vspace{-5mm}
\end{figure}

Several architectures using ReRAM have been proposed for CIM-based AI~ \cite{ankit2019puma}. These accelerators use analog computing on ReRAM crossbars by utilizing Kirchhoff’s law (Fig.~\ref{fig:intro}a). While this approach promises power and latency benefits, in practice, many architectures are limited by (1) inefficient analog-to-digital conversion or vice versa, and (2) the low resilience to faults in analog computing. The former is caused by ADC/DAC drawing large amounts of power~\cite{ankit2019puma}. The latter results from stored values being altered by write and read accesses~\cite{liu2021fault,staudigl2021neurohammer}.

\begin{figure*}[t]
    \centering
    \includegraphics[width=.89\textwidth]{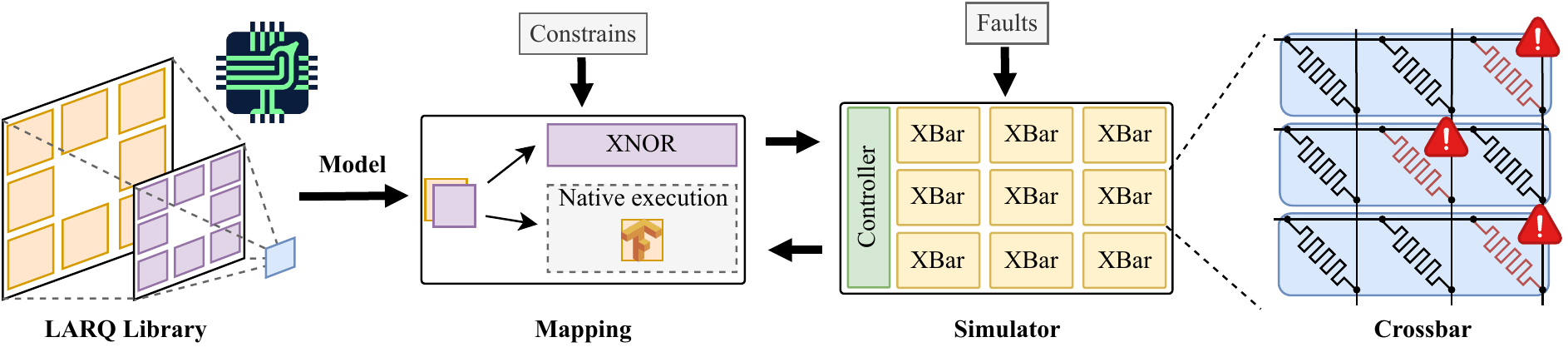}
    \caption{\faultyVP's end-to-end simulation workflow.}
    \label{fig:workflow}
    \vspace{-5mm}
\end{figure*}

Logic-in-memory~\cite{gaillardon2016programmable} is an orthogonal method to analog CIM that does not suffer from the mentioned limitations. It uses binary vectors stored in crossbar columns. Two columns are combined with given logic operations, and the result is stored in a third column (Fig.~\ref{fig:intro}b)~\cite{kvatinsky2014magic}. This method is less error-prone, as only two states (high and low resistive) are stored, and expensive ADCs/DACs are not required~\cite{gaillardon2016programmable}. Despite worse latency and density compared to analog CIM, LIM effectively accelerates AI applications in memory~\cite{papandroulidakis2017crossbar}. Nevertheless, LIM still suffers from faults caused by immature technology~\cite{liu2021fault}, including both traditional and unique faults.
Traditional faults are represented by conventional fault models prevalent in CMOS-based memories. Unique faults are emerging faults associated with memristive devices~\cite{8624895}. Despite their impact on individual memristors, faults might be acceptable for AI workloads if accuracy is not significantly reduced. This effect is well understood for analog computing~\cite{rasch2021flexible} but not for LIM.

\textbf{Contributions}: We present the first thorough investigation of the impact of \textit{traditional faults} on binary logic families.
Thereby, we measure the resilience of binary logic families based on two introduced metrics. Furthermore, we propose \faultyVP, an end-to-end mapping and simulation framework for binary neural networks (BNNs) using LIM in ReRAM.

\section{Background}\label{sec:back}
\subsection{Related Work}
Previous work investigated non-idealities in memristive crossbars and their impact on CIM~\cite{7017558,7116247}. Various simulation frameworks have been proposed to investigate the impact of faults on machine learning applications. Chakraborty et al.~\cite{9218688} presented a generalized approach to simulate neural networks on faulty memristive crossbars. The framework can simulate linear and non-linear non-idealities at architectural level. He et al.~\cite{8806787} proposed an end-to-end neural network tool that builds upon PyTorch. The tool takes into account the non-ideal effects of crossbars and adjusts mapping and training to drastically limit the impact of errors. The existing research has focused on analog-based CIM without exploiting binary logic families mapped to the memristive crossbar. 

\subsection{Binary Neural Networks (BNNs)}
BNNs emerged as a promising low-power, low-cost, and reduced accuracy approach using aggressive quantization~\cite{qin2020binary}. This method is particularly promising to deploy deep neural networks to resource-constrained devices, such as on the edge. The accuracy of BNNs is not on par with its less-quantized counterparts. This limitation is an unsolved challenge for today’s complex datasets. However, simple classification tasks can achieve competitive performance. Table 2 in~\cite{qin2020binary} reports up to 98.4\% accuracy for the MNIST dataset using an MLP, but accuracy drops to 40\% for Imagenet using VGG. Due to the low area and power cost of BNNs, an ecosystem of commercial tools recently emerged. For example, the Larq library~\cite{larq} provides reference implementations and functions for training and deploying BNNs. In a BNN, the basic arithmetic scheme of convolutions in neural networks—the matrix-matrix-multiplication—is equivalent to an XNOR operation between two single-bit precision vectors. This arithmetic relation maps directly to LIM for memristive crossbars. Hence, a BNN is currently a preferred operation mode for neuromorphic devices using ReRAM in edge applications. While convolutional and dense layers can be represented as XNOR operations, and therefore executed on ReRAMs, this is not the case for other, non-binary operations. We decided to take a conservative approach, in which other operations (e.g., integer bit-count operations after each layer) are executed in CMOS in the hardware model.

\subsection{LIM families on Memristive Crossbars}
\begin{figure*}[ht]
    \centering
    \hfill \subfloat[]{
        \includegraphics[width=\columnwidth]{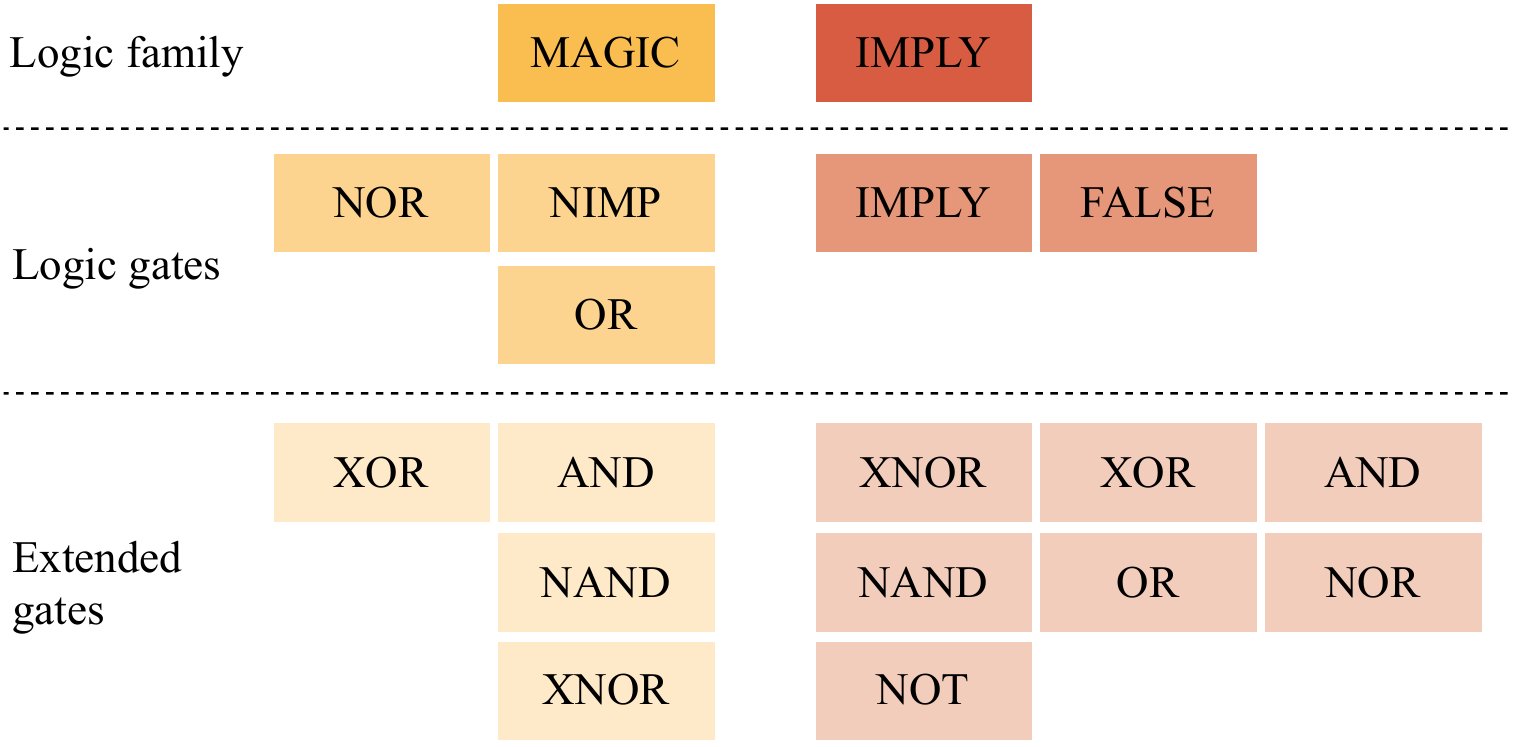}
        \label{fig:operations}
    }    \quad
    \hfill \subfloat[]{\adjustbox{width=0.75\columnwidth,valign=B,raise=0\baselineskip}{%
        \begin{tabular}{|c|c|c|c|c|}
        \cline{2-5}\multicolumn{1}{r|}{} & \multicolumn{2}{c|}{\textbf{IMPLY}} & \multicolumn{2}{c|}{\textbf{MAGIC}} \\
        \cline{2-5}\multicolumn{1}{r|}{} & \multicolumn{1}{l|}{\textbf{\# mem}} & \multicolumn{1}{l|}{\textbf{\# cycle}} & \multicolumn{1}{l|}{\textbf{\# mem}} & \multicolumn{1}{l|}{\textbf{\# cycle}} \\
        \hline
        \textbf{AND} & 3 & 4 & 5 & 9 \\
        \hline
        \textbf{IMP} & 2 & 1 & / & / \\
        \hline
        \textbf{NAND} & 3 & 3 & 5 & 12 \\
        \hline
        \textbf{NOR} & 3 & 5 & 3 & 1 \\
        \hline
        \textbf{NOT} & 2 & 2 & / & / \\
        \hline
        \textbf{OR} & 3 & 3 & 3 & 1 \\
        \hline
        \textbf{XNOR} & 4 & 6 & 4 & 6 \\
        \hline
        \textbf{NIMP} & / & / & 3 & 1 \\
        \hline
        \textbf{XOR} & 4 & 5 & 3 & 3 \\
        \hline
        \end{tabular}%
      }
      \label{tab:logic_fam_internals}
      }
    \caption{Simulation methodology: (a) overview of basic operations and the extended logic gates, and (b) number of memristive devices and required clock cycles of the implemented logic families.}
    \hfill
\vspace{-8mm}
\end{figure*}

As stated in the introduction, memristive crossbars can be used in an analog or LIM fashion. We investigate LIM as it trades higher error resilience and reduced ADC/DAC overhead with lower power density and higher latency. In LIM, the logical state (0 or 1) is represented as a high or low resistive value programmed in a memristor. A set of memristors are operating together for any logical operation to build a certain logic gate. An operation voltage is applied for a logical operation between two inputs. It is modulated by the state of the input memristors and applied to the output memristor. This voltage alters the state of the output memristor. Logic families have been classified into three categories: statefulness, proximity of computation, and flexibility~\cite{8106959}. Within the scope of this work, we focus on MAGIC~\cite{kvatinsky2014magic} and IMPLY~\cite{kvatinsky2013memristor}, which define basic logic operations. A full set of operations can be defined by daisy-chaining the basic ones. Fig.~\ref{fig:operations} illustrates the basic operations (OPs) supported by the logic families. To compute the inference of a BNN, we extend the basic OPs towards more sophisticated OPs, including the dominant XNOR operation.

\section{Simulation Methodology}
\label{sec:methodology}
\faultyVP's end-to-end workflow is presented in Fig.~\ref{fig:workflow}. The input to the flow is a user-defined BNN model in the Larq library~\cite{larq} in Tensorflow.
\faultyVP~iterates the BNN, maps XNOR-compatible operations to the memristive crossbar simulator, and executes the remainder of the model in native Tensorflow (representing CMOS logic next to a crossbar).

\textbf{Mapping}: \faultyVP's mapping tool brings convolutional and dense layers of a BNN to a crossbar of a given size. The mapper takes the kernel values from a BNN Tensorflow model and generates crossbar write, read, and logic instructions. It tracks if partial kernels fit on the remaining places to minimize write operations. We apply a linear mapping that iterates the BNN layerwise. The mapper writes all instructions as a binary file containing memory addresses of kernels and values, plus required logic operations.

\textbf{Simulator}: The simulator implements a memory controller that parses the given binary file from the mapping tool and provides the parameterized crossbars with the kernels and input values. The controller executes the logic operations of the logic families. The simulator tracks the resulting pulses applied to the respective bit and word lines for each crossbar access (read, write, and compute operation). The tracking information is used to trigger the currently active fault models.

\textbf{Crossbar}: The crossbar module implements a memristive crossbar array including a set of fault models. We focused on the conventional fault models, which are detailed in~\cite{8624895,9497692}. The crossbar consists of memristors with binary states that connect the respective bit with word lines. Each memristor can be assigned with one of the following fault models:

\begin{itemize}
\item \textbf{Stuck-at-Fault (SAF)} is modeled as a constant resistive value of the memristor, which is either the high resistive state (HRS) or the low resistive state (LRS).
\item \textbf{Read-Distructive-Fault (RDF)} flips the current state of the memristive cell and returns a correct value.
\item \textbf{Deceptive Read Destructive Fault (DRDF)} alters the current state of the cell but returns the incorrect value.
\item \textbf{Incorrect Read Fault (IRF)} does not change the cell state but returns an incorrect value.
\item \textbf{Slow Write Fault (SWF)} does not successfully write the cell, and hence returns the unmodified value.
\end{itemize}

The fault models are randomly assigned to a certain percentage of all instantiated memristors, defined as the \textit{injection rate}. To better understand the resilience of faults on the functional behavior of logic families, we introduce two new metrics. The \textit{Quality of Logic (QoL)} is defined for a single fault model as 
\begin{equation}
    \text{QoL} = \sum_{i=0}^{G-1} \frac{ \Lambda }{ \Omega } \cdot 100\%,
\end{equation}
where G is the number of gate types, $\Lambda$ the total number of faulty outputs, and $\Omega$ the number of all outputs. QoL implies how well the entire set of supported logic gates performs when affected by a certain fault. The \textit{Impact of Fault (IoF)} is defined for a single gate type as 
\begin{equation}
    \text{IoF} = \sum_{i=0}^{F-1} \frac{ \Lambda }{ \Omega } \cdot 100\%,
\end{equation}
where F is the number of fault types. IoF indicates the impact of all faults on a single logic gate. These two metrics are calculated on the basis of the information provided in Table~\ref{tab:logic_fam_internals}, which shows the required number of memristors and the resulting cycle count of each logic gate.

\begin{figure*}[ht]
    \centering
    \subfloat[]{
        % IMPLY
        \begin{tabular}{|c|c|c|c|c|c}
        \cline{2-6}\multicolumn{1}{r|}{} &
          \multicolumn{1}{l|}{\textbf{SAF}} &
          \multicolumn{1}{l|}{\textbf{RDF}} &
          \multicolumn{1}{l|}{\textbf{DRDF}} &
          \multicolumn{1}{l|}{\textbf{IRF}} &
          \multicolumn{1}{l|}{\textbf{IoF}}
          \\
        \hline
        \textbf{AND} &
          \cellcolor[rgb]{ 1,  .851,  .4}42\% &
          \cellcolor[rgb]{ 1,  .851,  .4}42\% &
          \cellcolor[rgb]{ .922,  .835,  .392}39\% &
          \cellcolor[rgb]{ 1,  .851,  .4}42\% &
          \multicolumn{1}{c|}{\textbf{38\%}}
          \\
        \hline
        \textbf{IMP} &
          \cellcolor[rgb]{ .898,  .831,  .388}38\% &
          \cellcolor[rgb]{ .6,  .784,  .365}25\% &
          \cellcolor[rgb]{ 1,  .733,  .345}50\% &
          \cellcolor[rgb]{ 1,  .733,  .345}50\% &
          \multicolumn{1}{c|}{\textbf{34\%}}
          \\
        \hline
        \textbf{NAND} &
          \cellcolor[rgb]{ .796,  .816,  .38}33\% &
          \cellcolor[rgb]{ .6,  .784,  .365}25\% &
          \cellcolor[rgb]{ 1,  .608,  .286}58\% &
          \cellcolor[rgb]{ 1,  .851,  .4}42\% &
          \multicolumn{1}{c|}{\textbf{34\%}}
          \\
        \hline
        \textbf{NOR} &
          \cellcolor[rgb]{ 1,  .851,  .4}42\% &
          \cellcolor[rgb]{ 1,  .851,  .4}42\% &
          \cellcolor[rgb]{ .796,  .816,  .38}33\% &
          \cellcolor[rgb]{ 1,  .851,  .4}42\% &
          \multicolumn{1}{c|}{\textbf{37\%}}
          \\
        \hline
        \textbf{NOT} &
          \cellcolor[rgb]{ 1,  .733,  .345}50\% &
          \cellcolor[rgb]{ 1,  .733,  .345}50\% &
          \cellcolor[rgb]{ 1,  .365,  .173}75\% &
          \cellcolor[rgb]{ 1,  .733,  .345}50\% &
          \multicolumn{1}{c|}{\textbf{48\%}}
          \\
        \hline
        \textbf{OR} &
          \cellcolor[rgb]{ .796,  .816,  .38}33\% &
          \cellcolor[rgb]{ .6,  .784,  .365}25\% &
          \cellcolor[rgb]{ 1,  .851,  .4}42\% &
          \cellcolor[rgb]{ 1,  .851,  .4}42\% &
          \multicolumn{1}{c|}{\textbf{31\%}}
          \\
        \hline
        \textbf{XNOR} &
          \cellcolor[rgb]{ 1,  .824,  .388}44\% &
          \cellcolor[rgb]{ .898,  .831,  .388}38\% &
          \cellcolor[rgb]{ 1,  .824,  .388}44\% &
          \cellcolor[rgb]{ 1,  .733,  .345}50\% &
          \multicolumn{1}{c|}{\textbf{41\%}}
          \\
        \hline
        \textbf{QoL} &
          \textbf{40\%} &
          \textbf{35\%} &
          \textbf{49\%} &
          \textbf{45\%} &
          
          \\
        \cline{1-5}\end{tabular}%
        \label{tab:res_fam_imply}
    } \hfil \subfloat[]{
                % MAGIC
        \begin{tabular}{|c|c|c|c|c|c}
        \cline{2-6}\multicolumn{1}{r|}{} &
          \multicolumn{1}{l|}{\textbf{SAF}} &
          \multicolumn{1}{l|}{\textbf{RDF}} &
          \multicolumn{1}{l|}{\textbf{DRDF}} &
          \multicolumn{1}{l|}{\textbf{IRF}} &
          \multicolumn{1}{l|}{\textbf{IoF}}
          \\
        \hline
        \textbf{AND} &
          \cellcolor[rgb]{ .882,  .831,  .388}35\% &
          \cellcolor[rgb]{ .631,  .788,  .365}25\% &
          \cellcolor[rgb]{ .882,  .831,  .388}35\% &
          \cellcolor[rgb]{ 1,  .812,  .384}43\% &
          \multicolumn{1}{c|}{\textbf{31\%}}
          \\
        \hline
        \textbf{NIMP} &
          \cellcolor[rgb]{ .839,  .824,  .384}33\% &
          \cellcolor[rgb]{ .733,  .808,  .376}29\% &
          \cellcolor[rgb]{ 1,  .471,  .224}67\% &
          \cellcolor[rgb]{ .945,  .839,  .392}38\% &
          \multicolumn{1}{c|}{\textbf{34\%}}
          \\
        \hline
        \textbf{NAND} &
          \cellcolor[rgb]{ 1,  .776,  .365}45\% &
          \cellcolor[rgb]{ .694,  .8,  .373}28\% &
          \cellcolor[rgb]{ 1,  .565,  .267}60\% &
          \cellcolor[rgb]{ 1,  .565,  .267}60\% &
          \multicolumn{1}{c|}{\textbf{45\%}}
          \\
        \hline
        \textbf{NOR} &
          \cellcolor[rgb]{ 1,  .796,  .373}44\% &
          \cellcolor[rgb]{ .945,  .839,  .392}38\% &
          \cellcolor[rgb]{ .839,  .824,  .384}33\% &
          \cellcolor[rgb]{ 1,  .824,  .388}42\% &
          \multicolumn{1}{c|}{\textbf{36\%}}
          \\
        \hline
        \textbf{XOR} &
          \cellcolor[rgb]{ 1,  .706,  .333}50\% &
          \cellcolor[rgb]{ 1,  .353,  .169}75\% &
          \cellcolor[rgb]{ 1,  .706,  .333}50\% &
          \cellcolor[rgb]{ .945,  .839,  .392}38\% &
          \multicolumn{1}{c|}{\textbf{46\%}}
          \\
        \hline
        \textbf{OR} &
          \cellcolor[rgb]{ .867,  .827,  .388}34\% &
          \cellcolor[rgb]{ .682,  .8,  .373}27\% &
          \cellcolor[rgb]{ 1,  .851,  .4}40\% &
          \cellcolor[rgb]{ 1,  .851,  .4}40\% &
          \multicolumn{1}{c|}{\textbf{30\%}}
          \\
        \hline
        \textbf{XNOR} &
          \cellcolor[rgb]{ 1,  .796,  .373}44\% &
          \cellcolor[rgb]{ .945,  .839,  .392}38\% &
          \cellcolor[rgb]{ 1,  .796,  .373}44\% &
          \cellcolor[rgb]{ 1,  .706,  .333}50\% &
          \multicolumn{1}{c|}{\textbf{40\%}}
          \\
        \hline
        \textbf{QoL} &
          \textbf{41\%} &
          \textbf{37\%} &
          \textbf{47\%} &
          \textbf{44\%} &
          \\
        \cline{1-5}\end{tabular}%
        \label{tab:res_fam_magic}
    }\hfil  \subfloat[]{
    \begin{tikzpicture}
    % IMPLY
        \pgfplotstableread[row sep=\\, col sep=&]{
        fault & fault2 & stuck-at & RDF & IRF & DRDF & SWFset & SWFreset & AVG\\
        1  & 5  & 100.00 & 75.00 & 50.00 & 73.68 & 26 & 0  & 74.67\\
        3  & 10 & 28.00  & 41.00 & 36.00 & 30.00 & 26 & 20 & 33.75\\
        5  & 15 & 27.00  & 30.00 & 16.00 & 16.00 & 20 & 7  & 22.25\\
        8  & 20 & 36.84  & 22.00 & 20.00 & 19.00 & 12 & 15 & 24.46\\
        10 & 25 & 14.00  & 11.00 & 21.00 & 19.00 & 20 & 7  & 16.25\\
        12 & 30 & 15.00  & 12.63 & 18.00 & 20.00 & 16 & 17 & 16.41\\
        15 & 35 & 10.53  & 8.00  & 19.00 & 18.00 &    &    & 13.88\\
        20 & 40 & 14.00  & 6.00  & 16.84 & 18.00 &    &    & 13.71\\
        25 & 45 & 5.00   & 4.00  & 15.00 & 20.00 &    &    & 11.00\\
        30 & 50 & 13.00  & 7.00  & 18.00 & 20.00 &    &    & 14.50\\
        }\dataset
        \begin{axis}[
            ylabel={Accuracy (\%)},
            xlabel={Injection rate (\%)},
            ylabel style={at={(0.018, 0.5)}, anchor=north},
            legend style={at={(0.5,1.25)}, anchor=north, legend cell align={left}},
            %legend image post style={scale=1},
            legend columns=4,
            %%% If space on the right, choose this one:
            %legend style={at={(0.82,0.96)}, anchor=north, legend cell align={center}},
            %legend columns=1,
            width=\columnwidth,
            ymajorgrids = true,
            xmajorgrids = true,
            height=5cm,
            ymin=0,
            ymax=110,
            ytick={10, 20, 30, 40, 50, 60, 70, 80, 90, 100},
            xtick={1, 5, 10, 15, 20, 25, 30},
            ylabel style ={font=\small},
            xlabel style ={font=\small},
            tick label style={font=\small}
            ]
            \addplot[draw=cg1, thick] table[x=fault, y=stuck-at]{\dataset};
            \addplot[draw=cg2, thick, loosely dotted] table[x=fault, y=DRDF]{\dataset};
            \addplot[draw=cg3, thick, densely dashed] table[x=fault, y=IRF]{\dataset};
            \addplot[draw=cg4, thick, dashdotted] table[x=fault, y=RDF]{\dataset};
            \addplot[draw=cg5, thick, densely dotted] table[x=fault2, y=SWFreset]{\dataset};
            \addplot[draw=cg6, thick, loosely dashed] table[x=fault2, y=SWFset]{\dataset};
            \addplot[draw=red, thick, mark=square*,mark options={fill=red}] table[x=fault, y=AVG]{\dataset};  
            \legend{SAF, DRDF, IRF, RDF, SWFreset, SWFset, AVG}
        \end{axis}
    \end{tikzpicture}
    \label{fig:fault_vs_acc_imply}
    } \hfil	\subfloat[]{
    \begin{tikzpicture}
    % MAGIC
        \pgfplotstableread[row sep=\\, col sep=&]{
        fault & fault2 & stuck-at & IRF & DRDF & RDF & SWFreset & SWFset & AVG\\
        1  & 5  & 62.22 & 84.00 & 84.00 & 87.00 & 59 & 35 & 79.31  \\
        3  & 10 & 37.78 & 53.00 & 60.00 & 62.22 & 21 & 13 & 53.25  \\
        5  & 15 & 5.00  & 26.00 & 50.00 & 19.00 & 16 & 15 & 25.00  \\
        8  & 20 & 30.00 & 21.00 & 16.84 & 20.00 & 9  & 13 & 21.96  \\
        10 & 25 & 17.00 & 19.00 & 18.00 & 20.00 & 13 & 14 & 18.50  \\
        12 & 30 & 27.37 & 15.00 & 17.89 & 20.00 & 3  & 15 & 20.07  \\
        15 & 35 & 20.00 & 23.00 & 14.00 & 20.00 &    &    & 19.25  \\
        20 & 40 & 19.00 & 15.00 & 14.00 & 19.00 &    &    & 16.75  \\
        25 & 45 & 11.00 & 17.00 & 9.00  & 18.00 &    &    & 13.75  \\
        30 & 50 & 9.00  & 19.00 & 8.00  & 19.00 &    &    & 13.75  \\
        }\dataset
        \begin{axis}[
            ylabel={Accuracy (\%)},
            xlabel={Injection rate (\%)},
            ylabel style={at={(0.018, 0.5)}, anchor=north},
            legend style={at={(0.5,1.25)}, anchor=north, legend cell align={left}},
            legend columns=4,
            %%% If space on the right, choose this one:
            %legend style={at={(0.82,0.96)}, anchor=north, legend cell align={center}},
            %legend columns=1,
            width=\columnwidth,
            ymajorgrids = true,
            xmajorgrids = true,
            height=5cm,
            ymin=0,
            ymax=110,
            ytick={10, 20, 30, 40, 50, 60, 70, 80, 90, 100},
            xtick={1, 5, 10, 15, 20, 25, 30},
            ylabel style ={font=\small},
            xlabel style ={font=\small},
            tick label style={font=\small}
            ]
            \addplot[draw=cg1, thick] table[x=fault, y=stuck-at]{\dataset};
            \addplot[draw=cg2, thick, loosely dotted] table[x=fault, y=DRDF]{\dataset};
            \addplot[draw=cg3, thick, densely dashed] table[x=fault, y=IRF]{\dataset};
            \addplot[draw=cg4, thick, dashdotted] table[x=fault, y=RDF]{\dataset};
            \addplot[draw=cg5, thick, densely dotted] table[x=fault2, y=SWFreset]{\dataset};
            \addplot[draw=cg6, thick, loosely dashed] table[x=fault2, y=SWFset]{\dataset};
            \addplot[draw=red, thick, mark=square*,mark options={fill=red}] table[x=fault, y=AVG]{\dataset};  
            \legend{SAF, DRDF, IRF, RDF, SWFreset, SWFset, AVG}
        \end{axis}
    \end{tikzpicture}
    \label{fig:fault_vs_acc_magic}
    }
    \caption{Simulation results: Impact of faults on (a) the IMPLY logic family and (b) the MAGIC logic family. Effect of injection rate on the BNN inference accuracy for (c) IMPLY and (d) MAGIC.}
\vspace{-4mm}
\end{figure*}
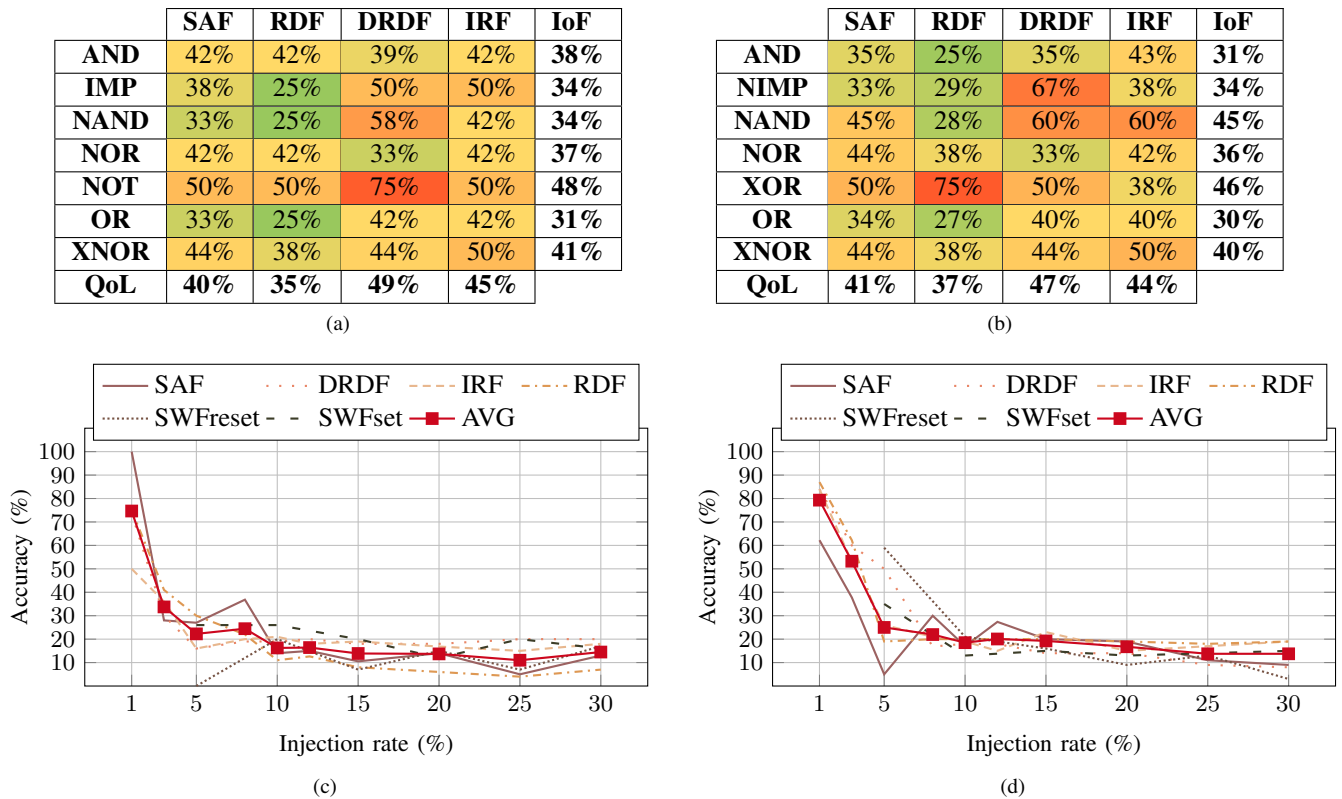

\section{Results and Discussion}
\label{sec:res}
In this section, we validate the simulation framework and investigate the robustness of two available logic families with respect to a sub-set of traditional faults. Furthermore, we investigate the fault impact on the accuracy of a BNN.

Table~\ref{tab:res_fam_imply} and Table~\ref{tab:res_fam_magic} illustrate the simulation results of logic gates versus the injected faulty behavior. The analysis is based on the execution of single operations of all possible input values and initial values of all memristive devices. As an example, the NOT gate of the IMPLY logic family uses two distinct memristors (Table~\ref{tab:logic_fam_internals}). The memristors are initialized with the patterns [(0,0), (1,0), (0,1), (1,1)], and the inputs 0 and 1 are supplied. The shown percentages represent the number of wrong outputs and are visualized with a heat map, which indicates that a reduced number of faults was propagated at logic level, affecting the functional behavior of the logic family. Note that coupling faults have not been considered in this first set of experiments because they require two or more consecutive accesses. Both QoL and IoF are calculated and shown in the last row and column of both tables, respectively. The OR gate performs the best in terms of resilience for both logic families, with an IoF of 31\% (IMPLY) and 30\% (MAGIC). The calculated QoL indicates that RDF has the least impact on logic gates, while the DRDF has a high impact. In addition, the worst logic gate in terms of fault resilience is the NOT (IMPLY) and the XOR gate (MAGIC). In general, \textit{the experiment shows that both logic families perform equally well considering the similar QoL and IoF values}. However, architectural design decisions can be optimized based on the resilience towards certain fault models. For instance, the designer should favor the IMPLY implementation of the NAND gate with an IoF of 34\% over the MAGIC implementation with an IoF of 45\%.

Next, we perform a preliminary experiment with the full simulation framework to investigate the impact on a BNN during inference. We use a BNN model from the LARQ library examples~\cite{larq} and train it with the MNIST dataset~\cite{lecun1998gradient}. Within the scope of this work, only a preliminary simulation was performed, using a subset of the test data set for the inference. Each configuration has been executed twenty times to eliminate the effects resulting from the randomly placed errors based on the given injection rate. Fig.~\ref{fig:fault_vs_acc_imply} and Fig.~\ref{fig:fault_vs_acc_magic} illustrate the impact of the injected faults on the accuracy of the BNN. Independent of the chosen logic family, it can be observed that already \textit{low injection rates significantly reduce the accuracy of the BNN}. Furthermore, the accuracy reaches a plateau at around 15\%, independent of the logic family. We observe that the SWFset and SWFreset reduce the accuracy to roughly 20\% for the MAGIC version, independent of the injection rate.

\section{Conclusion}
\label{sec:con}
This work investigated the impact of a sub-set of traditional faults on the logic-in-memory paradigm in general and on the accuracy of BNNs in particular. We developed an end-to-end simulation framework that maps the XNOR operations of an arbitrary Tensorflow model onto memristive crossbar arrays. In addition, the crossbar includes different faults randomly injected into the memristive devices. The results show that faults on memristive crossbar arrays significantly impact their functionality. Furthermore, a comparison of two logic families based on their fault resilience is facilitated by introducing two novel metrics. In future work, we plan to run extensive simulations with different models and data sets. In addition, we intend to expand the framework by including memristor-specific faults as well as the implementation of other logic families.

%\vspace{-4mm}
\bibliographystyle{IEEEtran}
\bibliography{bibliography.bib}

\end{document}